\documentclass[12pt]{iopart}

\usepackage{graphicx}
\usepackage[breaklinks=true,colorlinks=true,linkcolor=blue,urlcolor=blue,citecolor=blue]{hyperref}

\expandafter\let\csname equation*\endcsname\relax
\expandafter\let\csname endequation*\endcsname\relax

\usepackage{amsmath}
\usepackage{mathrsfs}

\DeclareMathOperator{\sign}{sign}

\begin{document}

\title[Breaking the Warp Barrier]{Breaking the Warp Barrier:
Hyper-Fast Solitons in Einstein-Maxwell-Plasma Theory
}

\author{Erik W. Lentz }

\address{Institut f\"ur Astrophysik, Georg-August Universitat G\"ottingen, 
                 G\"ottingen, Germany 37077}
\ead{erik.lentz@uni-goettingen.de}
\vspace{10pt}
\begin{indented}
\item[] \today
\end{indented}

\begin{abstract}
Solitons in space--time capable of transporting time-like observers at superluminal speeds have long been tied to violations of the weak, strong, and dominant energy conditions of general relativity. The negative-energy sources required for these solitons must be created through energy-intensive uncertainty principle processes as no such classical source is known in particle physics. This paper overcomes this barrier by constructing a class of soliton solutions that are capable of superluminal motion and sourced by purely positive energy densities. The solitons are also shown to be capable of being sourced from the stress-energy of a conducting plasma and classical electromagnetic fields. This is the first example of hyper-fast solitons resulting from known and familiar sources, reopening the discussion of superluminal mechanisms rooted in conventional physics.

\end{abstract}

\section{Introduction}

Hyper-fast solitons within modern theories of gravity have been a topic of energetic speculation in recent decades \cite{Alcubierre1994,Everett1996,Pfenning1997,Hiscock1997,Krasnikov1998,Olum1998,VanDenBroeck1999,Millis1999,Visser2000,Loup2001,Natario2002,Gauthier2002,Lobo2003,Lobo2004,Lobo2007,Obousy2008,White2013,DeBenedictis2018}. One of the most prominent critiques of compact mechanisms of superluminal motion within general relativity is that the geometry must largely be sourced from a form of negative energy density, though there are no such known macroscopic sources in particle physics. Other concerns include difficulties associated with constructing a soliton from a nearly flat space--time up to the superluminal phase, where the transported central observers become surrounded by a horizon, and the equal difficulties of evolving from the superluminal phase back the flat space--time. Challenges associated with creating horizons also include communication between inside and outside observers through the soliton shell, bombardment of the inside observers by Hawking radiation, and stress-energy buildup on the leading horizon. Further, creating a self-sustaining Alcubierre-type superluminal soliton \cite{Alcubierre1994} of $100$~m radius would need an immense amount of (magnitude) energy, in excess of the scale that is in the visible universe, $E_{tot} \sim - 6 \times 10^{62} v_s/c$~kg mass equivalent \cite{Pfenning1997}, though some progress has been made in this area, reducing the required energy to $\sim -10^{30} v_s/c$~kg mass equivalent \cite{VanDenBroeck1999,Obousy2008}, and even down to the kilogram and gram scale \cite{Krasnikov2003}.

This paper addresses the first critique by constructing a new class of hyper-fast soliton solutions within general relativity that are sourced purely from positive energy densities, thus removing the need for exotic negative-energy-density sources. This is made possible through considering hyperbolic relations between components of the space--time metric's shift vector, which depart from the elliptic or linear relations that limited solitons in the previous literature to require negative energies. Further, the stress-energy sourcing these solutions fits the form of a classical electronic plasma, placing superluminal phenomena into the purview of known physics. The remainder of the paper is structured as follows: Section~\ref{GRTheory} presents the geometry of these novel solitons using the ADM formalism \cite{ADM} and produces the components of the Einstein equation relevant for the class of solutions; Section~\ref{PosE} introduces the conditions of the hyperbolically-related shift vectors and the rules for constructing a class of solutions with everywhere-positive energy density and conventional energy-momentum conditions and demonstrates these qualities for a family of solutions; Section~\ref{GREM} solves the dynamical component of the geometry via the Einstein equation trace and derives requirements for a potential sourcing plasma; and Section~\ref{Conclusions} discusses the consequences of discovering a superluminal mechanism driven by known sources and potential avenues for future study.

\section{Solitons in General Relativity}
\label{GRTheory}

The space--times considered here are decomposed in the ``3+1'' (ADM) formalism using a similar convention to that presented in \cite{MTW}, or \cite{BARDEEN1983}, specifically following the latter's sign protocol. The line element of the space--time is cast in the form
\begin{equation}
    ds^2 = -\left(N^2-N^i N_i \right) dt^2 - 2 N_i dx^i dt + h_{ij} dx^i dx^j,
\end{equation}
where the time coordinate $t$ stratifies space--time into space-like hypersurfaces, the space metric components $h_{ij}$ evaluated at $t$ provide the intrinsic geometry of that hypersurface, and the similarly-evaluated shift vector components $N^i$ at $t$ provide the coordinate three-velocity of the hypersurface's normal. The time-like unit normal one-form is therefore proportional to the coordinate time element $\mathbf{n}^* = N dt$, and the unit normal vector $\mathbf{n}$ to the hypersurface has components
\begin{equation}
    n^{\nu} = \left(\frac{1}{N}, \frac{N^i}{N}  \right).
\end{equation}
Einstein summation notation is used throughout this paper, with Greek indices running over space--time components and Latin indices over space components. The lowering of Latin indices is performed using the hypersurface metric $h$ unless otherwise stated. Natural units $G=c=1$ are used. Lastly, the lapse function $N$ is set to unity.

Central to the computation of the Einstein tensor is the hypersurface extrinsic curvature, which can be written as the negative covariant derivative of the normal vector field $\mathbf{n}$, or in terms of coordinate derivatives as
\begin{equation}
    K_{ij} = -\frac{1}{2} \left( \partial_t h_{ij} + N^k \partial_k h_{ij} + \partial_i N^k h_{kj} + \partial_j N^k h_{ki} \right).
\end{equation}
The solutions considered here will have hypersurfaces parameterized by flat metrics under Cartesian coordinates $h_{ij} = \delta_{ij}$, reducing the extrinsic curvature expression to the symmetric combination of shift vector derivatives. The trivial form of $N$ and $h$ imply that the Eulerian observers, time-like observers whose motion in space--time is normal to the hypersurfaces with four-velocity $\mathbf{n}$, are in free fall.

Resolving the behavior of solitons within general relativity begins with a check of the weak energy condition and the momentum conditions. The weak energy condition is given by the projection of the Einstein equation onto the hypersurface normal
\begin{equation}
    G^{\mu \nu} n_{\mu} n_{\nu} = \left( R^{\mu \nu} - \frac{1}{2} g^{\mu \nu} R \right)n_{\mu} n_{\nu}= 8 \pi T^{\mu \nu} n_{\mu} n_{\nu},
\end{equation}
where the projected stress-energy is to be called the local Eulerian energy density
\begin{equation}
    T^{\mu \nu} n_{\mu} n_{\nu} = N^2 T^{00} = E.
\end{equation}
The geometric side of the energy constraint equation is divisible into the intrinsic hypersurface curvature $^{(3)}R$, and the extrinsic curvature's trace $K = K^i_i$ and its quadratic hypersurface scalar $K^i_j K^j_i$
\begin{equation}
    8 \pi E = \frac{1}{2}\left( ^{(3)}R - K^i_j K^j_i + K^2 \right).
\end{equation}
The contribution of the hypersurface intrinsic curvature vanishes as the space metric $h$ is flat. The purely geometric portion of the energy condition may then be expanded in terms of the shift vector components 
\begin{align}
    K^2 - K^i_j K^j_i  &= 2 \partial_x N_x \partial_y N_y + 2 \partial_x N_x \partial_z N_z + 2 \partial_z N_z \partial_y N_y \nonumber \\
    &- \frac{1}{2} \left( \partial_x N_y + \partial_y N_x  \right)^2 - \frac{1}{2} \left( \partial_x N_z + \partial_z N_x  \right)^2 - \frac{1}{2} \left( \partial_z N_y + \partial_y N_z  \right)^2. \label{Egeom}
\end{align}
Note that the last three elements of the above expression are negative definite, while the first three are of indeterminant type. These first three terms have the potential to provide the energy function with an island of configurations that satisfy the weak energy condition. The first task of this work will be to show there exist non-flat compact moving configurations that have everywhere non-negative energy.

The momentum conditions are implemented here by comparing the mixed projection local Eulerian momentum density,
\begin{equation}
    J_i = - n_{\alpha} T^{\alpha}_i = N T^0_i,
\end{equation}
to the mixed projection of the Einstein tensor, resulting for the considered geometries in the three conditions
\begin{equation}
    8 \pi J_i = \partial_j K^j_i - \partial_i K.
\end{equation}
Both the energy and momentum conditions must be satisfied everywhere and will provide a sense for the stress-energy sources needed to construct the soliton geometries.

The dynamics of the geometry are in general set by the remaining six free components of the Einstein equation. Several of these degrees have already been made moot by the choice of  a flat $h$ and constant lapse function $N$. The conditions for positive energy solutions introduced in the next section will reduce the number of dynamical geometric degrees of freedom to one, meaning that only a single component of the dynamical portion of the Einstein equation is needed. The trace condition is a natural choice, given by
\begin{equation}
    8 \pi T^{\mu}_{\mu} = - R, \label{trace}
\end{equation}
where the space--time Ricci scalar decomposes in this class of space--times to
\begin{equation}
    R = K^2 + K^i_j K^j_i + 2 \mathscr{L}_{\mathbf{n}} K, \label{RicciScalar}
\end{equation}
where $\mathscr{L}_{\mathbf{n}}()$ is the Lie derivative in the direction of the normal unit vector field.

\section{Constructing Positive-Energy Solutions Using a Hyperbolic Shift Vector Potential}
\label{PosE}

The class of geometries studied here will be characterized by a shift vector potential function, a real-valued function $\phi$ with spatial gradient relating the shift vector components
\begin{equation}
    N_i = \partial_i \phi.
\end{equation}
The soliton potentials considered here will be set to a steady state, moving with constant velocity and allowing the potential to be parameterized by displacement from its moving center $\phi(x-x_s(t),y-y_s(t),z-z_s(t))$, where $\dot{x}_s(t)=v_x$, $\dot{y}_s(t)=v_y$, and $\dot{z}_s(t)=v_z$ are the constant velocity components of the soliton.

The potential condition alone is insufficient to produce a positive definite function of Eqn.~\ref{Egeom}, and so a relation between all the shift vector components is added. The most common relations explored in the literature are linear and elliptic. Specifically, the linear relation ($N_x = N_y = 0$) of Ref.~\cite{Alcubierre1994} produced the renowned toroid of negative energy density about the soliton bubble of $N_z$, here displayed in Cartesian coordinates,
\begin{equation}
    E_{\text{Alc}} = \frac{-1}{32 \pi} \left( \left( \partial_x N_z  \right)^2 + \left( \partial_y N_z \right)^2 \right).
\end{equation}
The expansionless ($K = -1/2(\partial_x N_x + \partial_y N_y + \partial_z N_z) = 0$) elliptic relation of Ref.~\cite{Natario2002} restricted the energy form to the negative definite square of the extrinsic curvature
\begin{equation}
    E_{\text{Nat}} = \frac{-1}{16 \pi} K^i_j K^j_i.
\end{equation}
Parabolic and hyperbolic relations remained to be explored.

The hyperbolic relation is examined here. Specifically, the potential function will be taken to satisfy a linear wave equation over the spatial coordinates
\begin{equation}
    \partial_x^2 \phi +\partial_y^2 \phi -\frac{2}{v_h^2} \partial_z^2 \phi = \rho,
\end{equation}
where $ v_h/\sqrt{2}$ is the dimensionless wave front `speed' on the hypersurface, and $\rho$ is the source function. The positive $z$-axis is singled out as it will be the principal direction of travel for the soliton. Therefore, the remainder of this paper will consider only motion along the $z$ direction, setting $v_x = v_y = 0$. The geometric side of the energy condition can then be rewritten as
\begin{equation}
    K^2 - K^i_j K^j_i  = 2 \partial_x^2 \phi \partial_y^2 \phi + 2 \partial_z^2 \phi \left( \frac{2}{v_h^2} \partial_z^2 \phi + \rho \right) - 2 \left(\partial_y \partial_x \phi  \right)^2 - 2 \left(\partial_z \partial_x \phi  \right)^2 - 2 \left(\partial_y \partial_z \phi  \right)^2.
\end{equation}

It is still not altogether clear what the sign of the energy function is, so two simplifications are applied for the purpose of demonstration. Assuming that $\rho$ and $\phi$ are both parameterized in the ($x,y$) coordinates by the $l_1$ norm $s= |x| + |y|$, the energy further can be further simplified to a two-coordinate form, here using $(z,x)$,
\begin{equation}
    E  = \frac{1}{16 \pi} \left(2 \partial_z^2 \phi \left(\rho + \frac{2}{v_h^2} \partial_z^2 \phi \right) - 4 \left( \partial_z \partial_x \phi \right)^2 \right). \label{Edens}
\end{equation}
The Green's function representation of the potential, holding that the potential's initial condition at $z \to - \infty$ is null, takes the form
\begin{equation}
    \phi = \int d x' dz' \frac{1}{4 v_h} \Theta \left( z-z' - \frac{|\Delta x|}{v_h} \right) \rho (z',|x'| + |y|),
\end{equation}
where $\Theta()$ is the Heaviside function and $\Delta x = x-x'$. The shift vectors can then be found in the Green's form
\begin{align}
    N_z &= \frac{1}{4 v_h} \int d x'  \rho \left(z- \frac{|\Delta x|}{v_h}, |x'| + |y| \right), \\
    N_x &= - \frac{1}{4 v_h^2} \int d x' \sign \left( \Delta x \right) \rho \left(z- \frac{|\Delta x|}{v_h}, |x'| + |y| \right),
\end{align}
where $\sign \left( \right)$ is the sign function. One can see that the shift vector components are proportional to integrals of source over the `past' wave cone. Given the Green's expressions, it can be straightforwardly computed that $|\partial_z^2 \phi| \ge v_h |\partial_z\partial_x \phi|$, implying that the energy condition satisfies the inequality
\begin{align}
    E &\ge 2  \rho \times \partial_z^2 \phi \nonumber \\
    &= \rho \times\frac{1}{ 2 v_h} \int dx' \partial_r \rho(r,|x'| + |y|)|_{r= z-|\Delta x|/v_h} ,
\end{align}
from which rules may be formed to ensure the energy density is everywhere non-negative. For instance, the energy function will be non-negative for configurations such that the local source density and the $z$-component source density gradient integrated along the intersecting `past' wave trajectories are of the same sign. 

\begin{figure}
\begin{center}
\includegraphics[height=9cm]{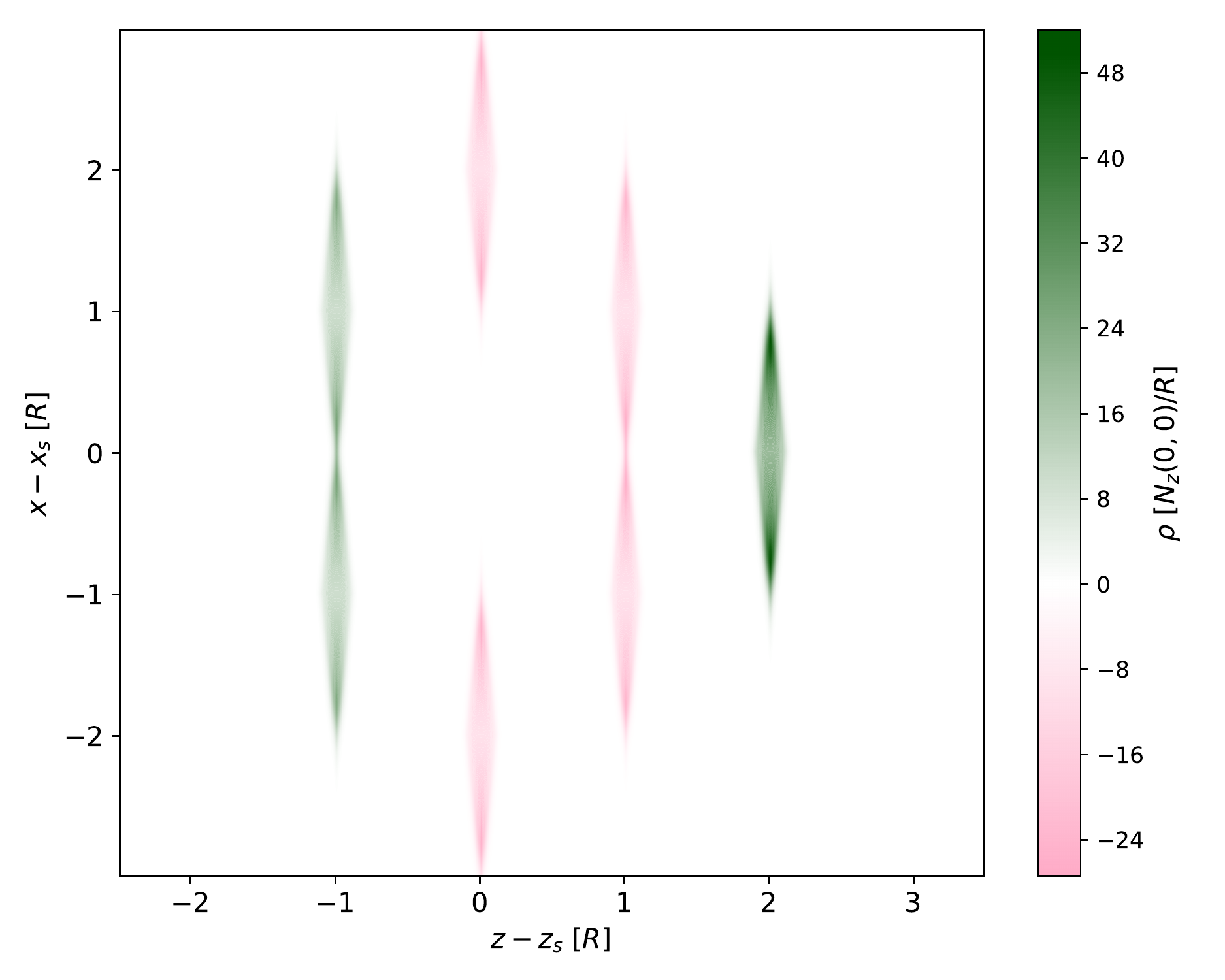}
\caption{Projection of the source $\rho$ of shift vector potential $\phi$ along $(x,0,z)$. Propagation direction of the soliton is along the $z$-axis. Charge within each chord perpendicular to the long axis of the sources are calibrated to give a level surface in the central region. Shape and charge profile of each rhomboid source are identical. Total integrated charge of the system is 0.}
\label{fig:rho}
\end{center}
\end{figure}

\begin{figure}
\begin{center}
\includegraphics[height=9cm]{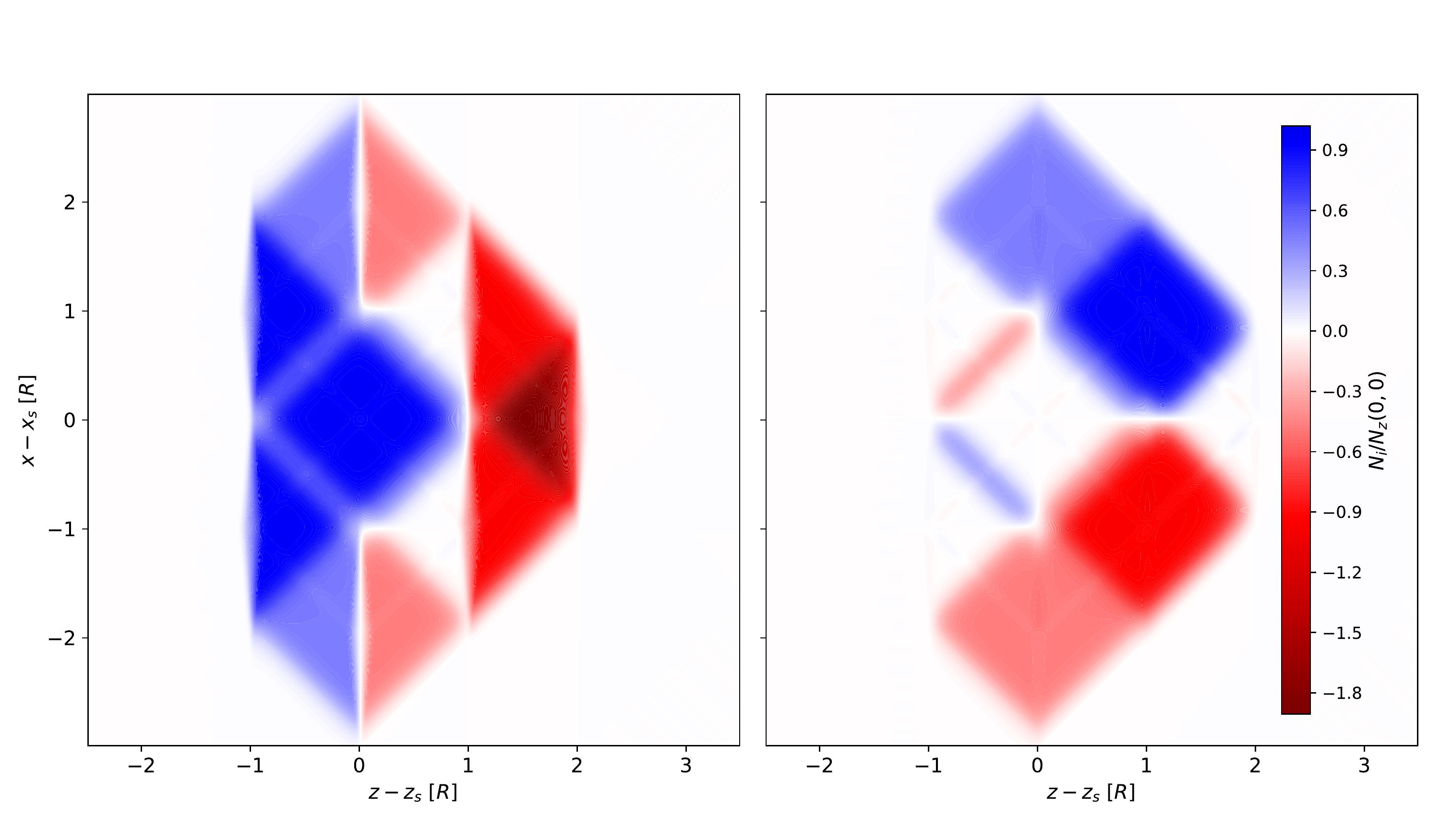}
\caption{Projection of the shift vector components $N_z$ (left) and $N_x$ (right) along $(x,0,z)$. Propagation direction of the soliton is from left to right along the $z$-axis. The multi-compartment structure is a distinct departure from the single top-hat soliton fond in \cite{Alcubierre1994} and \cite{Natario2002}. Total integrated shift in each direction is 0.}
\label{fig:Ns}
\end{center}
\end{figure}

Consider the pentagonal configuration of sources in Fig.~\ref{fig:rho}, illustrated via bi-lateral $s$-projection onto the $(x,0,z)$ plane of a hypersurface, as a demonstration of one such compact positive energy configuration with net motion $v_z = v_s$. The configuration is such that the spatial wave fronts traveling from the left-most beams create a broad region of high and level $N_z$ at the center, terminating on the right-most pair of beams of opposing density, with the remaining sources organized to terminate the stray branches of the wave cone, Fig.~\ref{fig:Ns}. The net hyperbolic source of the soliton is zero. The individual sources are formed as rhomboids in the 2D projection such that the boundary lines are angled to be between the trajectories of the hyperbolic wavefront cone and the $z$-constant plane for the purpose of satisfying the ``non-negative energy rule'' above. The perpendicular components of the shift vector are seen to vanish in the central region, while the parallel component over the same region is also very level but non-zero. The placid region at the soliton center is nearly tidal-force-free, where Eulerian observers move along essentially straight lines at $v_{rel} = N_z(0)-v_o$ relative to the soliton. This is in contrast to the volatile boundary where there exist domains in which the shift vector can be much greater in size and divergent in direction. The net shift vector of the soliton is found to be zero.

The relation between the soliton velocity is assigned to be  consistent with the shift vector of the soliton's central region ($v_s = N_z(0,0)$) as Eulerian observers in the central region will then travel along time-like curves with proper time rate matching those far from the soliton, $d \tau = d t$. In this case, the logistics of soliton travel for observers in the central region reflect those in \cite{Alcubierre1994}.

\begin{figure}
\begin{center}
\includegraphics[height=9cm]{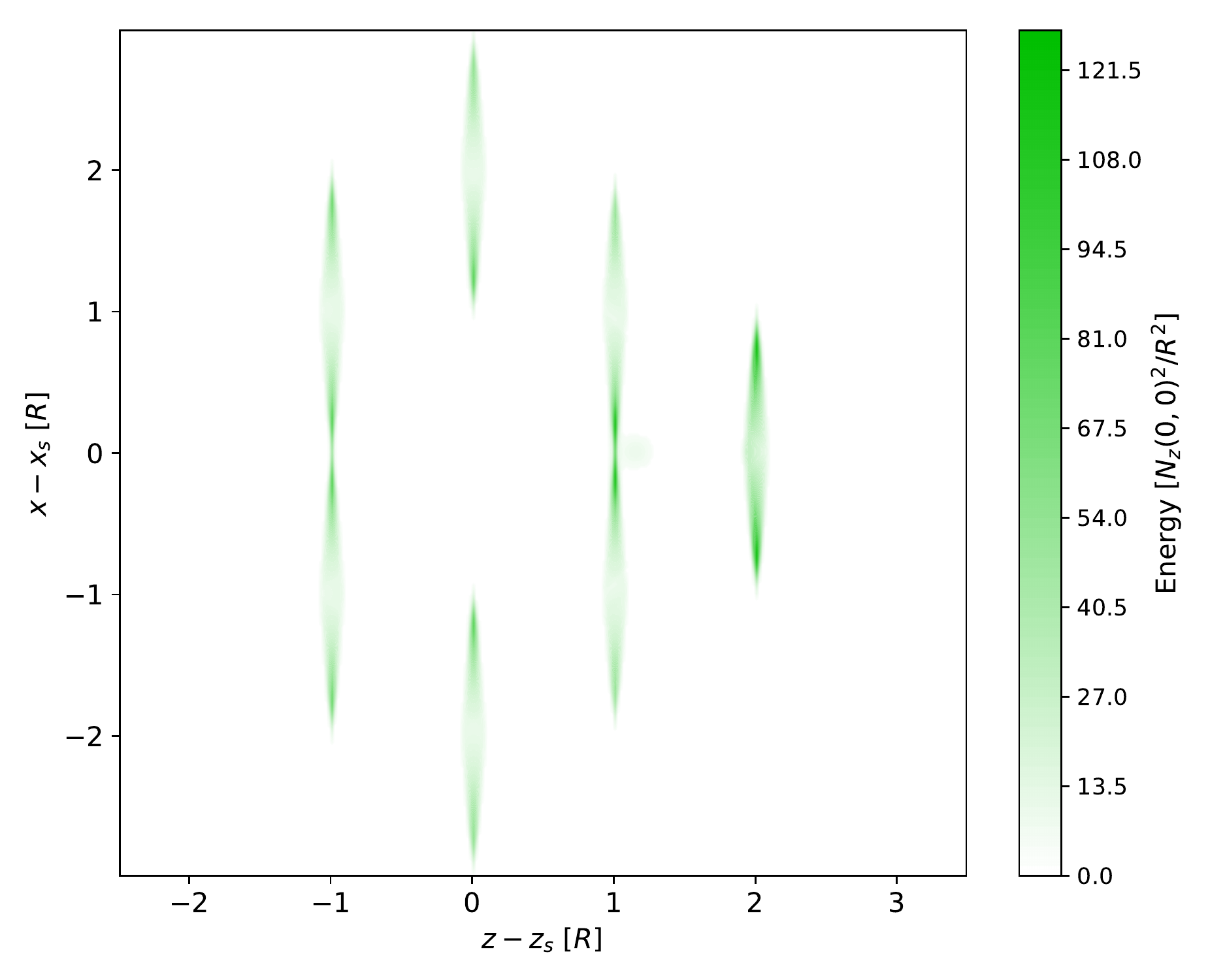}
\caption{Projection of the local energy density of Eqn.~\ref{Edens} along $(x,0,z)$. Propagation direction of the soliton is from left to right along the $z$-axis. The energy density is dominated by those regions containing hyperbolic source $\rho$, but also extends weakly to the boundaries of the wavefronts. The energy density is everywhere positive and therefore satisfies the weak energy condition.}
\label{fig:energy}
\end{center}
\end{figure}

The energy density of the soliton is seen to be positive definite in Fig.~\ref{fig:energy}. Each rhomboid source $\rho_{\text{rhom}}$ is constructed individually to have everywhere positive energy density and to have positive energy density in the presence of other sources of $\rho_{\text{rhom}}$ of the same size and orientation. One can therefore piece together many other solutions from these elements of hyperbolic source. The total energy requirements of the positive-energy solitons closely follow that of Ref.~\cite{Pfenning1997} as applied to the Alcubierre solution
\begin{equation}
    E_{tot} = \int E \sqrt{-g} d^3x.
\end{equation}
For solitons where the radial extent of the central region $R$ is much larger than the thickness of the energy-density laden boundary shell $w$ ($w \ll R$), the energy is estimated to be
\begin{equation}
    E_{tot} \sim  C v_s^2 \frac{R^2}{w}
\end{equation}
where $C$ is a form factor typically of order unity. The required energy for a positive-energy soliton with central regions radius $R = 100$~m and average source thickness along the z-axis $w = 1$~m approaches a mass equivalent of $E_{tot} \sim (\text{few}) \times 10^{-1} M_{\odot} v_s$, which is of the same magnitude as the estimate of Ref.~\cite{Pfenning1997} for an Alcubierre solution of the same dimensions, but without the uncertainties associated with where one might source the energy. The estimate for the Alcubierre solution sourced by naturally occurring Casimir forces is much higher, $\sim -6 \times 10^{62} v_s$~kg, which requires one to reduce the boundary thickness to a few hundred Planck lengths. However, no such naturality conditions are known to restrict the stress-energy driving the positive energy solutions. Further, many soliton solutions have been made since Ref.~\cite{Alcubierre1994} that drastically improved on the overall negative energy requirements \cite{VanDenBroeck1999,Loup2001,Krasnikov2003,Obousy2008,White2013}. Several of these approaches may provide significant savings in energy for the positive-energy soliton.

\begin{figure}
\begin{center}
\includegraphics[height=9cm]{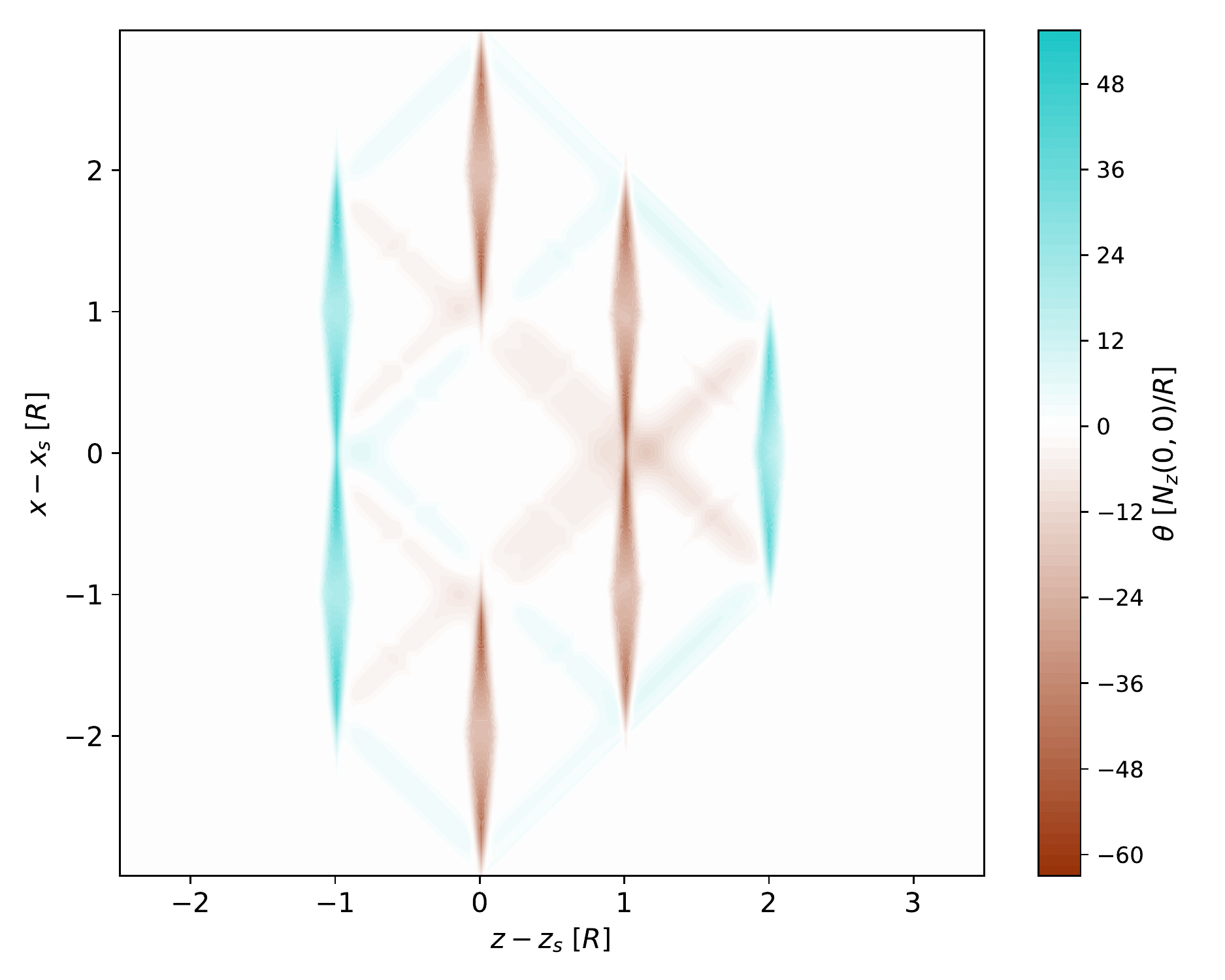}
\caption{Projection of the local volume expansion factor $\theta$ along $(x,0,z)$. Propagation direction of the soliton is from left to right along the $z$-axis. Positive and negative expansion factor are largely associated with negative and positive hyperbolic sources respectively. Non-zero expansion factor also exist in the spaces in-between hyperbolic sources along the hyperbolic wavefronts. Total integrated expansion factor is 0.}
\label{fig:thetaexp}
\end{center}
\end{figure}

The hypersurface volume expansion, calculated here from the extrinsic curvature trace $\theta = K$, can be found in Fig.~\ref{fig:thetaexp}. The volume expansion of the positive-energy soliton is complex, containing expansions and contractions on all sides of the central region. The solution of \cite{Alcubierre1994} possesses only one negative expansion lobe at the leading edge of its soliton and one positive expansion lobe at the trailing edge. The largest values of $\theta$ for the positive-energy soliton coincide with the sources of stress-energy density, in contrast to the solution of Ref.~\cite{Alcubierre1994} where the energy density and expansion factor are maximally separated on the soliton boundary. Further, the largest positive and negative lobes of $\theta$ on the positive-energy soliton are seen to correlate to the negative and positive hyperbolic sources respectively. Both solutions have net expansion of 0.

The momentum conditions under the hyperbolic shift vector potential are seen to vanish
\begin{equation}
    J_i = 0,
\end{equation}
implying a trivial net energy--momentum relation relative to free-falling Eulerian observers. Sources with multiple species can can satisfy zero net momentum flux while each variety is non-static. The multi-species stress-energy source of choice in this paper is an electric plasma consisting of a massive fluid and electromagnetic fields, the conditions of which are investigated further in the next section.

\section{Soliton-Plasma Dynamics}
\label{GREM}

\begin{figure}
\begin{center}
\includegraphics[height=9cm]{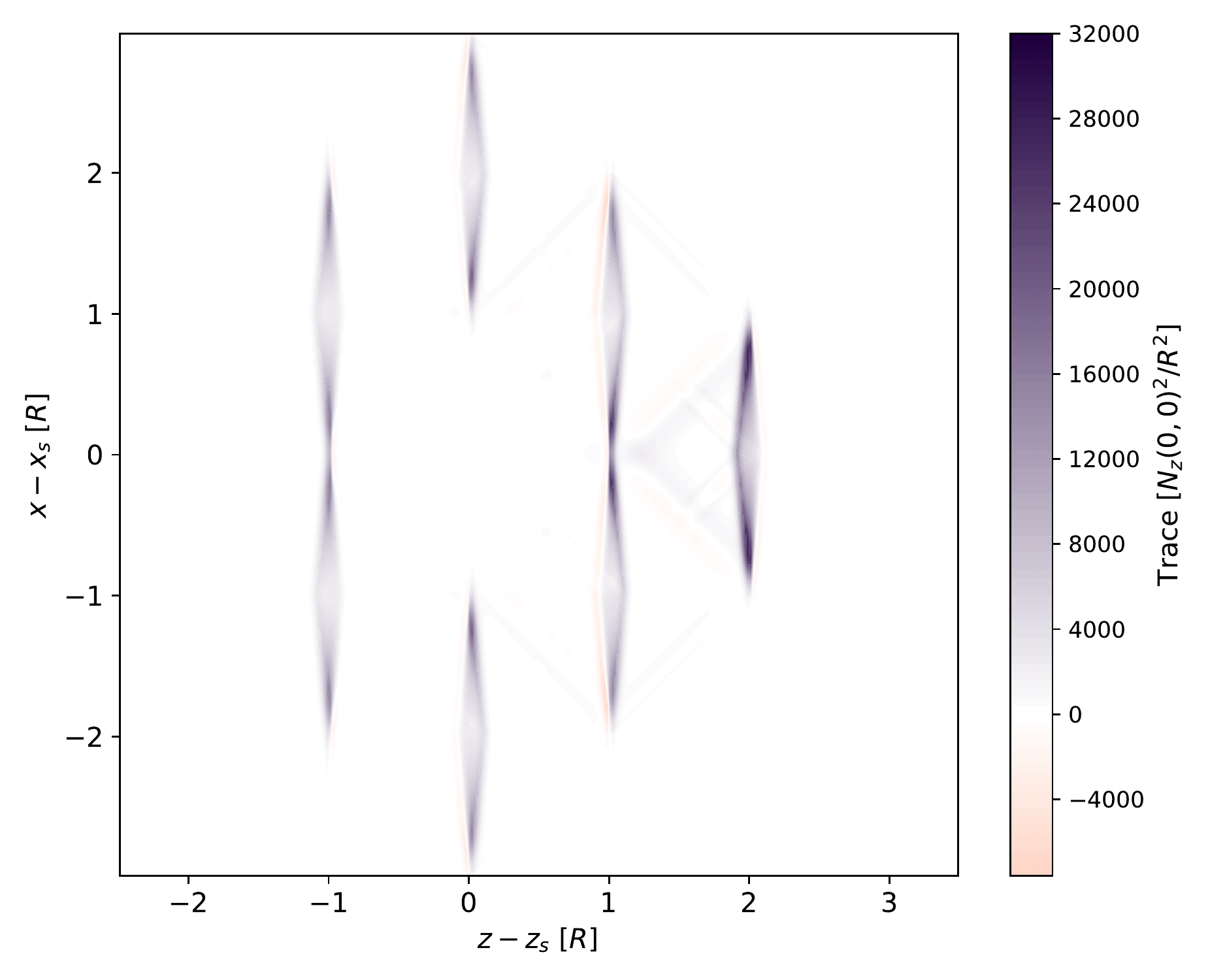}
\caption{Projection of the Einstein tensor trace from Eqn.~\ref{traceEqn} along $(x,0,z)$. Propagation of the soliton is taken to be uniform with speed consistent with shift vector in the central region, $v_s = N_z(0,0)$.}
\label{fig:trace}
\end{center}
\end{figure}

This section describes the conditions needed for an electrically conducting plasma to act as source for the positive-energy soliton. The dynamics of the hyperbolic potential $\phi$, or equivalently the hyperbolic source $\rho$, can be set by the Einstein equation trace, Eqn.~\ref{trace}. The Ricci scalar of Eqn.~\ref{RicciScalar} under the conditions of the previous section becomes
\begin{equation}
    R = -16 \pi E + 2 \theta^2 + 16 \pi \left( \left(N_z - v_s \right) \partial_z K + 2 N_x \partial_x K \right).
\end{equation}
The stress-energy of the plasma plus electromagnetic fields is of the form
\begin{equation}
    T^{\mu \nu} = \left(\rho_m + p \right)u^{\mu} u^{\nu} + p g^{\mu \nu} + F^{\mu \alpha} F^{\nu \beta} g_{\alpha \beta} - \frac{1}{4} g^{\mu \nu} F^{\alpha \beta} F_{\alpha \beta},
\end{equation}
where $\rho_m$ is the plasma mass density, $p$ is the plasma pressure, $u^{\alpha}$ are the components of the plasma velocity field, and $F^{\mu \nu}$ are the components of the anti-symmetric field strength tensor. The trace condition then becomes
\begin{equation}
    -16 \pi E + 2 \theta^2 + 16 \pi \left( \left(N_z - v_s \right) J_z + 2 N_x J_x \right) = 8 \pi \left(\rho_m - 3p \right), \label{traceEqn}
\end{equation}
which on the stress-energy side involves only the massive fluid as the electromagnetic stress-energy is trace-less. Further, note that the energy and momentum conditions involve the plasma, the electromagnetic fields, as well as the shift vector
\begin{align}
    E &= \left(\rho_m + p\right) \left(u^0\right)^2 - p + \frac{1}{2} \left(1 + 2 N_i N^i \right) E^i E_i - \frac{1}{2} \left( N_i E^i \right)^2 + \frac{1}{2} B_i B^i - \frac{1}{2} \epsilon_{ijk} N^i E^j B^k, \\
    J_i &= \left(\rho_m + p \right) u_i u^0 - p N_i + \epsilon_{ijk} E^j B^k -  E_i \left(E_j N^j \right) \nonumber \\
    &- \frac{1}{2} N_i \left( \left(1 - N_k N^k \right) E_l E^l +  \left( N_k E^k \right)^2 - B_k B^k + \epsilon_{ljk} N^l E^j B^k \right),
\end{align}
where $E^i$ are the components of the electric field three-vector and $B^i$ are the components of the magnetic field pseudo-three-vector. The trace equation is used here to investigate the response of the fluid's mass and pressure density as the soliton velocity has already been set. In their trace combination, $\rho_m - 3 p$ can take on both positive and negative values, limited by the fluid equation of state. The solitons considered here, where the velocity matches the central shift vector ($v_s = N_z(0,0)$) has trace that is consistent with a fluid with equation of state $p \le \rho$, Fig.~\ref{fig:trace}, which is within the physically accepted range \cite{Neilsen2000}.

In addition to supporting the energy, momentum, and trace conditions for steady-state motion, the plasma must satisfy its own conditions. These include the Maxwell equations for the electric and magnetic fields, the conservation and dynamical equations for the massive component of the plasma, the pressure equation of state, and the additional relations between the massive and electric current densities. These conditions are of sufficient number to determine all the plasma's degrees, meaning that the geometric conditions cannot in general be used to dictate the state of media without over-constraining it, as there is only one geometric field and multiple independent geometric conditions. In addition to the energy and momentum conditions discussed above, causal contact is often used as a pre-condition for relativistic plasmas, and is frequently checked using the dominant energy condition. The dominant energy condition is respected by the sub-luminal solitons so long as the magnitude of the shift vector is less than unity in all domains ($N_i N^i < 1$). For higher speeds, the soliton begins to form horizons between its domains and the external vacuum. To further identify a solution of the more than dozen degrees of freedom of the plasma that satisfy the example soliton, and to investigate the horizon problems endemic to this and all other known superluminal solitons, would require computation beyond the scope of this paper. What can be said here is that the conditions of the plasma are consistent with the soliton geometry. It is now a matter of finding the right configuration.

\section{Conclusions}
\label{Conclusions}

This paper has demonstrated that there exist superluminal solitons in general relativity satisfying both the weak energy condition and the momentum conditions for conventional sources of stress-energy. This is the first known solution of its kind, as previous superluminal solitons have required large amounts of negative energy. The positive-energy geometries presented here distinguish themselves from the literature in that they obey a hyperbolic relation among their shift vector components in the form of a wave equation on the associated space-like hypersurface, whereas only linear or elliptic relations had been previously considered. The solitons were further constructed to contain a central region with minimal tidal forces, where proper time coincides with asymptotic coordinate time, and any Eulerian observer within the central region would remain stationary with respect to the soliton. The transport logistics of the presented positive-energy solitons are similar to the solitons of the Alcubierre solution. 

Beyond being the first positive energy solutions of their kind, the presented solitons appear to be a counter example to claims in the literature that superluminal space--times must violate the weak energy condition \cite{Olum1998,Lobo2003}. To contrast the present positive-energy soliton and the previous proofs, one must realize that the pre-conditions of these proofs are restricted to geometries with a single fastest causal path. No known generalizations of the proofs exist to extensive superluminal mechanisms. Previous superluminal solitons such as the Alcubierre and Nat\'ario solutions can be made to fit this condition as their structures permit a point-like limit of the interior, where the fastest causal path would settle. The example positive-energy, in contrast, must have non-trivial extent. Similar point-like limits of the positive energy solitons would collapse the geometry to the Minkowski vacuum, an expected outcome as the shift vector components are seen to integrate to zero. The positive energy soliton therefore appears to circumvent the criteria of \cite{Olum1998} and \cite{Lobo2003}.

The energy and momentum conditions of the presented positive-energy geometries were found to conform to a plasma with no net momentum flux. The trace of the Einstein equation, the single dynamical conditions that determines the hyperbolic shift vector potential, was used to determine the limits on the plasma equation of state in lieu of having already set the steady-state velocity of the soliton. The geometric conditions on the plasma are deferential to the plasma's own dynamical equations, which include equations of motion and constitutive relations for both the massive fluid and the electromagnetic fields. The total energy requirements of the positive-energy solitons appear to be of the same order as the original Alcubierre soliton under the same shell-thickness-to-diameter conditions, with the energy for a soliton of modest radius $R=100$~m and shell thickness $w=1$~m requiring $E_{tot} \sim (\text{few}) \times 10^{-1} M_{\odot} v_s/c$. This energy, though still immense, is intriguing as there have been many advances in reducing the required energy of the negative-energy solitons that may be equally effective for this new class of solutions. The next challenge is to bring the energy requirements of the positive-energy soliton to the human technological scale.

Once the energy requirement is lowered, the space--time signatures of positive-energy solitons may be studied in a laboratory setting using existing or novel methods. For instance, previous interferometric searches for hyperfast solitons could be recast to search for the much larger signal of a positive-energy-efficient soliton \cite{White2013,Lee2016}. The highly magnetized energetic and diffuse atmospheric plasma of magnetars may also be a natural place to look for signatures of positive-energy soliton geometries even prior to advances in energy reduction.

For theory, it is an appealing proposition to incorporate the degrees and dynamics of the plasma into the geometric computation. One could self-consistently simulate the creation, propagation, and dismantlement phases of a soliton at both sub- and superluminal speeds. Other directions include further optimizations of the solutions over the energy requirements and other trade-offs, the broadening of the soliton geometry to incorporate a ``payload'' in the soliton's central region, and studying the challenges of horizon formation when transitioning to superluminal speeds. However, developing models and configurations of the plasma alongside the geometry would in general require a large-scale numerical effort. Fortunately, in the era of gravitational-wave astronomy and high-precision cosmology, there exist a number of numerical relativity codes that are increasingly capable of describing massive fluids and gauge fields in relativistic space--time.

\section{Acknowledgements}

I would like to thank Justin Feng, Ken Olum, Jos\'e Nat\'ario, Jeffery Lee, Gerald Cleaver, Katy Clough, Leslie Rosenberg, Andreas Karch, and Sanjay Reddy for their insightful comments during the preparation of this manuscript.

\section*{References}
\bibliographystyle{iopart-num.bst}
\bibliography{Bibliography.bib}

\providecommand{\newblock}{}
\begin{thebibliography}{10}
\expandafter\ifx\csname url\endcsname\relax
  \def\url#1{{\tt #1}}\fi
\expandafter\ifx\csname urlprefix\endcsname\relax\def\urlprefix{URL }\fi
\providecommand{\eprint}[2][]{\url{#2}}

\bibitem{Alcubierre1994}
Alcubierre M 1994 {\em Classical and Quantum Gravity\/} {\bf 11} L73--L77
  \urlprefix\url{https://doi.org/10.1088%2F0264-9381%2F11%2F5%2F001}

\bibitem{Everett1996}
Everett A~E 1996 {\em Phys. Rev. D\/} {\bf 53}(12) 7365--7368
  \urlprefix\url{https://link.aps.org/doi/10.1103/PhysRevD.53.7365}

\bibitem{Pfenning1997}
Pfenning M~J and Ford L~H 1997 {\em Classical and Quantum Gravity\/} {\bf 14}
  1743--1751
  \urlprefix\url{https://doi.org/10.1088\%2F0264-9381\%2F14\%2F7\%2F011}

\bibitem{Hiscock1997}
Hiscock W~A 1997 {\em Classical and Quantum Gravity\/} {\bf 14} L183--L188
  \urlprefix\url{https://doi.org/10.1088\%2F0264-9381\%2F14\%2F11\%2F002}

\bibitem{Krasnikov1998}
Krasnikov S~V 1998 {\em Phys. Rev. D\/} {\bf 57}(8) 4760--4766
  \urlprefix\url{https://link.aps.org/doi/10.1103/PhysRevD.57.4760}

\bibitem{Olum1998}
Olum K~D 1998 {\em Phys. Rev. Lett.\/} {\bf 81}(17) 3567--3570
  \urlprefix\url{https://link.aps.org/doi/10.1103/PhysRevLett.81.3567}

\bibitem{VanDenBroeck1999}
Broeck C~V~D 1999 {\em Classical and Quantum Gravity\/} {\bf 16} 3973--3979
  \urlprefix\url{https://doi.org/10.1088%2F0264-9381%2F16%2F12%2F314}

\bibitem{Millis1999}
Millis M~G 1999 {\em Acta Astronautica\/} {\bf 44} 175 -- 182 ISSN 0094-5765
  missions to the Outer Solar System and Beyond
  \urlprefix\url{http://www.sciencedirect.com/science/article/pii/S0094576599000454}

\bibitem{Visser2000}
Visser M, Bassett B and Liberati S 2000 {\em Nuclear Physics B - Proceedings
  Supplements\/} {\bf 88} 267 -- 270 ISSN 0920-5632
  \urlprefix\url{http://www.sciencedirect.com/science/article/pii/S0920563200007829}

\bibitem{Loup2001}
{Loup} F, {Waite} D and {Halerewicz} E J 2001 {\em arXiv e-prints\/}
  gr-qc/0107097 (\textit{Preprint} \eprint{gr-qc/0107097})
  \urlprefix\url{http://www.jbis.org.uk/paper.php?p=2013.66.242}

\bibitem{Natario2002}
Nat{\'{a}}rio J 2002 {\em Classical and Quantum Gravity\/} {\bf 19} 1157--1165
  \urlprefix\url{https://doi.org/10.1088%2F0264-9381%2F19%2F6%2F308}

\bibitem{Gauthier2002}
{Gauthier} C, {Gravel} P and {Melanson} J 2002 {\em International Journal of
  Modern Physics A\/} {\bf 17} 2761

\bibitem{Lobo2003}
{Lobo} F and {Crawford} P 2003 {\em {Weak Energy Condition Violation and
  Superluminal Travel}\/} vol 617 p 277

\bibitem{Lobo2004}
Lobo F~S~N and Visser M 2004 {\em Classical and Quantum Gravity\/} {\bf 21}
  5871--5892
  \urlprefix\url{https://doi.org/10.1088\%2F0264-9381\%2F21\%2F24\%2F011}

\bibitem{Lobo2007}
{Lobo} F~S~N 2008 {\em Classical and Quantum Gravity Research\/}
  arXiv:0710.4474 (\textit{Preprint} \eprint{0710.4474})

\bibitem{Obousy2008}
{Obousy} R~K and {Cleaver} G 2008 {\em Journal of the British Interplanetary
  Society\/} {\bf 61} 364--369 (\textit{Preprint} \eprint{0712.1649})

\bibitem{White2013}
{White} H 2013 {\em Journal of the British Interplanetary Society\/} {\bf 66}
  242--247

\bibitem{DeBenedictis2018}
DeBenedictis A and Iliji{\'{c}} S 2018 {\em Classical and Quantum Gravity\/}
  {\bf 35} 215001 \urlprefix\url{https://doi.org/10.1088%2F1361-6382%2Faae326}

\bibitem{Krasnikov2003}
{Krasnikov} S 2003 {\em Phys. Rev. D\/} {\bf 67} 104013 (\textit{Preprint}
  \eprint{gr-qc/0207057})

\bibitem{ADM}
Arnowitt R, Deser S and Misner C~W 1959 {\em Phys. Rev.\/} {\bf 116}(5)
  1322--1330 \urlprefix\url{https://link.aps.org/doi/10.1103/PhysRev.116.1322}

\bibitem{MTW}
Misner C~W, Thorne K and Wheeler J 1973 {\em {Gravitation}\/} (San Francisco:
  W. H. Freeman) ISBN 978-0-7167-0344-0, 978-0-691-17779-3

\bibitem{BARDEEN1983}
Bardeen J~M and Piran T 1983 {\em Physics Reports\/} {\bf 96} 205 -- 250 ISSN
  0370-1573
  \urlprefix\url{http://www.sciencedirect.com/science/article/pii/0370157383900698}

\bibitem{Neilsen2000}
{Neilsen} D~W and {Choptuik} M~W 2000 {\em Classical and Quantum Gravity\/}
  {\bf 17} 733--759 (\textit{Preprint} \eprint{gr-qc/9904052})

\bibitem{Lee2016}
Lee J~S and Cleaver G~B 2016 {\em Physics Essays\/} {\bf 29} 201–206 ISSN
  0836-1398 \urlprefix\url{http://dx.doi.org/10.4006/0836-1398-29.2.201}

\end{thebibliography}

\end{document}